\documentclass[nolinenumbers]{aastex631}

\usepackage{xcolor}

\graphicspath{{./}{figures/}}

\begin{document}
\title{Detection Feasibility of H$_2$ in Ultra-hot Jupiter Atmospheres}

\author{Anastasia Morgan}
\affiliation{Laboratory for Atmospheric and Space Physics, University of Colorado, Boulder, CO 80303, USA}

\author{P. Wilson Cauley}
\affiliation{Laboratory for Atmospheric and Space Physics, University of Colorado, Boulder, CO 80303, USA}

\author{Kevin France}
\affiliation{Laboratory for Atmospheric and Space Physics, University of Colorado, Boulder, CO 80303, USA}

\author{Allison Youngblood}
\affiliation{Exoplanets and Stellar Astrophysics Lab, NASA Goddard Space Flight Center, Greenbelt, MD 20771, USA}

\author{Tommi T. Koskinen}
\affiliation{Lunar and Planetary Laboratory, University of Arizona, Tucson, AZ 85721-0092, USA}

\begin{abstract}
Ultra-hot Jupiters (UHJs) have recently been the focus of several atmospheric studies due to their extreme properties. While molecular hydrogen (H$_2$) plays a key role in UHJ atmospheres, it has not been directly detected on an exoplanet. To determine the feasibility of H$_2$ detection via transmission spectroscopy of the Lyman and Werner bands, we modeled UHJ atmospheres with H$_2$ rotational temperatures varying from 2000 K to 4000 K orbiting A-type stars ranging from $T_{eff}$ =  8,500 K to $T_{eff}$ = 10,300 K. We present simulated transmission spectra for each planet-star temperature combination while adding Poisson noise varying in magnitude from 0.5\% to 2.0\%. Finally, we cross-correlated the spectra with expected atmospheric H$_2$ absorption templates for each temperature combination. Our results suggest that H$_2$ detection with current facilities, namely the \textit{Hubble Space Telescope}, is not possible.  However, direct atmospheric transmission spectroscopy of H$_2$ may be viable with future UV-capable flagship missions.
\end{abstract}

\section{Introduction} \label{sec:intro}
UHJs are a fascinating population of exoplanets due to their extreme temperatures and resulting atmospheric chemistry, which more closely resembles late K-type stars than a typical planetary atmosphere. Molecular hydrogen plays a key role in heat transport in UHJ atmospheres as it dissociates and recombines across the terminator~\citep{bell2018}. Despite this and being the most abundant molecule in gas giant atmospheres~\citep{yelle2004aeronomy}, H$_2$ has not yet been directly detected in an exoplanet atmosphere. The large atmospheric scale heights of UHJs present optimal observational opportunities to detect H$_2$, particularly in the far-UV (FUV) where H$_2$ has strong rovibrational transitions in the Lyman and Werner bands. 

\section{Methods} \label{sec:style} 
To assess the potential detectability of H$_2$ in UHJ atmospheres, we simulated transmission spectra of planets around A-type host stars with a range of planet and stellar temperatures. Observation of planetary transmission spectra with H$_2$ transition lines between 1100 \AA\ and 1400 \AA\ requires a strong stellar FUV flux (the background light source) which can be provided by early-type host stars. We tested H$_2$ rotational temperatures ranging from 2,000 K to 4,000 K in intervals of 500 K and stellar temperatures from 8,500 K to 10,300 K in intervals of 300 K.

We fixed the planet radius at R$_p$ = 1.5 R$_{J}$ and assumed spherical symmetry for the atmosphere. For all planets we adopted a H$_2$ density profile derived for KELT-20 b, which was calculated using \texttt{GGchem}~\citep{woitke2021ggchem} and assumed an isothermal atmosphere for each temperature in the explored range. The isothermal assumption is reasonable given the small radial extent of the H$_2$ population. After using \texttt{GGchem} to calculate the mixing ratio of H$_2$ and the atmosphere's mean molecular weight as a function of pressure, we solved the hypsometric equation from \citet[][see their eq. 10]{koskinen2022mass} to determine the atmospheric height at each pressure. We then assumed a base number density of 10$^{17}$ cm$^{-3}$ at $r = 1.0 R_p$ and a pressure of 1 bar, where the H$_2$ mixing ratio is 0.848, and calculated the H$_2$ number density from $r = 1.0\ R_p$ out to $r = 1.3\ R_p$, where it drops to a negligible value of $\approx 10$ cm$^{-3}$. We note that KELT-20 b is near the lower range of our planetary temperatures and thus represents an \textit{optimistic} case for the hotter planets since H$_2$ dissociates at higher atmospheric pressures as the temperature increases, decreasing the depth of the H$_2$ transmission spectrum.

We determined H$_2$ rovibrational level populations using Maxwell-Boltzmann statistics, i.e., we assumed thermal equilibrium 
\citep{barthelemy07}. We then used the optical depth templates from \cite{mccandliss2003molecular} to calculate the relative optical depth of each rovibrational transition. Next, we scale the relative optical depth by the column density from our atmospheric model, resulting in an annulus of H$_2$ optical depth profiles around the planet.

We generated H$_2$ transmission spectra for each planet and stellar temperature combination at 10 equidistant transit times, which roughly mimics a 3-transit \textit{HST} program, where we determine the orbital parameters by assuming a circular orbit and solving for the semi-major axis necessary to give the assumed equilibrium temperature for the given host star $T_{eff}$. Finally, we add Poisson noise at the level of 0.5\%, 1.0\%, 1.5\%, and 2.0\% to each individual transmission spectrum. We note that we explored larger levels of Poisson noise but found that anything beyond $\approx 2\%$ required a very large number of simulated \textit{HST} exposures to reach a significant detection. 

For the cross-correlation analysis we first shifted the transmission spectra into the rest frame of the planet for assumed orbital velocities between 0 km s$^{-1}$ and 500 km s$^{-1}$. We averaged the resulting velocity-shifted transmission spectra and cross-correlated this average spectrum with a pre-computed H$_2$ absorption template for system velocities between $\pm$300 km s$^{-1}$. We restricted the cross-correlation calculation to between 1300 \AA\ and 1400 \AA\ where there is sufficient stellar flux and H$_2$ absorption lines to produce a signal. We repeated this experiment 500 times for each planet-star combination to test the robustness of the cross-correlation signal to the assumed noise levels. We took the average cross-correlation significance from these 500 trials as the final detection significance for each case. 

\section{Results} \label{sec:style}

\begin{figure}
    \centering
    \includegraphics[scale=0.367]{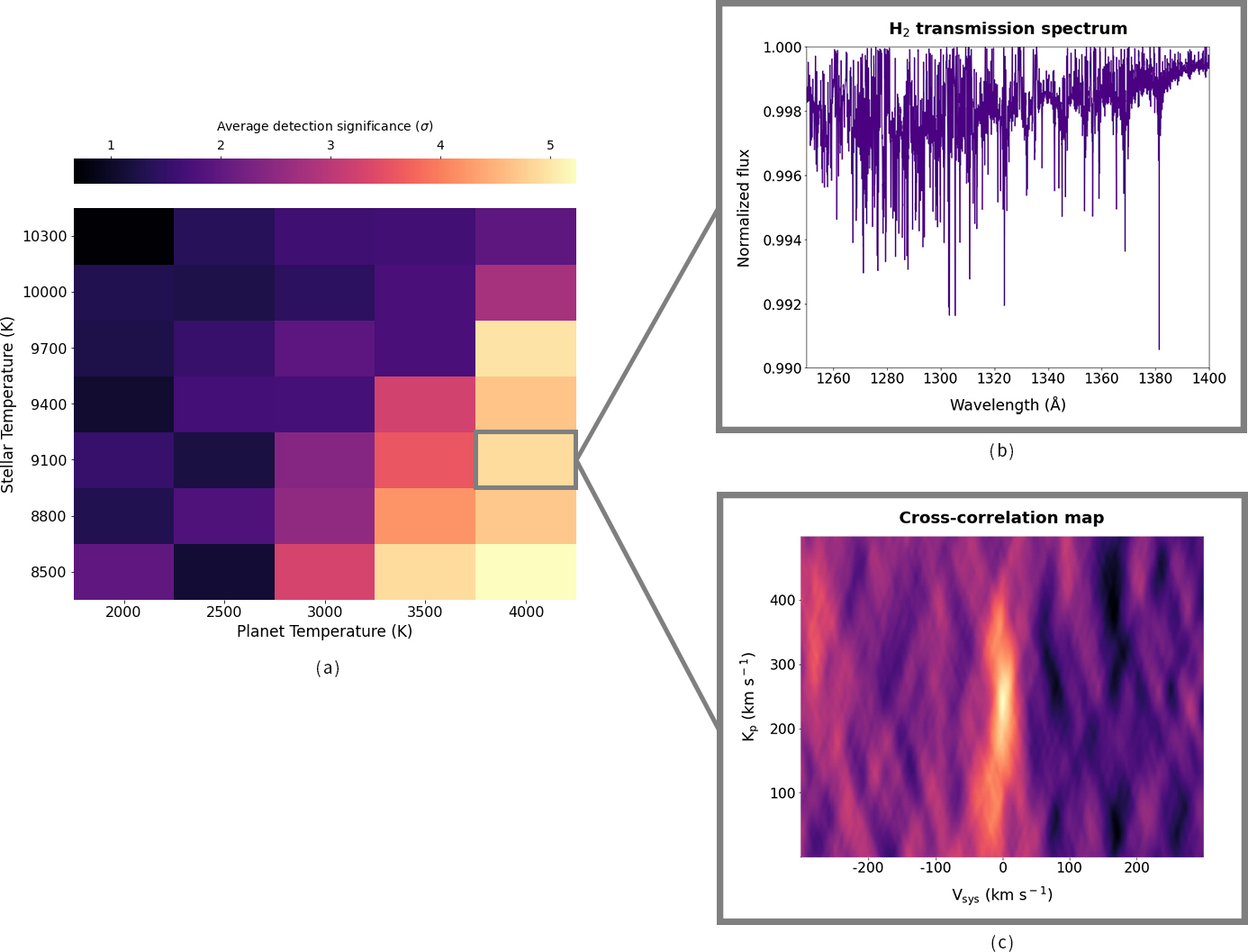}
    \caption{(a) Significance map for stellar and planet temperature combinations reflecting the average detection significance with 2\% transmission spectrum noise. (b) Noiseless transmission spectrum of H$_2$ example for $T_{planet}$ = 4000 K and $T_{eff}$ = 9100 K.  (c) Cross-correlation map example of $T_{planet}$ = 4000 K and $T_{eff}$ = 9100 K showing a signal detection of approximately 5$\sigma$.} 

\end{figure}

Our simulations suggest that high signal-to-noise is required for H$_2$ detection. As we show in Figure 1(c), H$_2$ is detectable with a signal significance of about 5$\sigma$ for a planet with $T_{planet} = 4000$ K orbiting a star with $T_{eff}$ = 9100 K and 2\% transmission spectrum noise. The signal map shown in Figure 1(a) indicates that the greatest detection of H$_2$ is approximately 5$\sigma$ and most likely to occur in higher temperature planets orbiting lower temperature stars for the considered temperature ranges. We note that this trend occurs due to the fixed planetary radius with increasing stellar radius; a fixed value of $R_p/R_*$ would likely produce better cross-correlation signals due to the increased signal-to-noise of the star's FUV flux. This trend is consistent with the lower transmission spectrum noise levels tested, where we measured signal detection significance values of $\approx12\sigma$ at 0.5\% noise, $\approx10\sigma$ at 1.0\% noise, and $\approx8\sigma$ at 1.5\%. 

\section{Conclusion} \label{sec:floats}
Given our assumed optimistic H$_2$ density profile, the simulated detection significance values are likely still unachievable with \textit{HST}. For example, using the STIS E140M exposure time calculator we estimate that the best signal-to-noise currently achievable for the FUV transmission spectrum of KELT-9 b, the hottest known UHJ orbiting a bright A0V star, is $\approx 3\%-5\%$. We estimate that a $5\sigma$ detection for KELT-9 b would require 6-7 transit observations of 5 \textit{HST} orbits each, or 30-35 total orbits. The $\leq 2\%$ noise levels required for detection via a reasonable amount of observing time will likely be achievable with next generation UV instruments, such as \textit{LUVOIR} and the LUMOS spectrograph~\citep{france2017}, although systematics may push the necessary transmission spectrum precision even lower. Finally, it is worth noting that our assumption of LTE for the H$_2$ gas in the lower atmosphere needs to be tested with models which account for the depth at which FUV photons are deposited in the atmosphere, potentially pumping the H$_2$ into non-thermal distributions. 

\bibliography{references}{}

\begin{thebibliography}{}
\expandafter\ifx\csname natexlab\endcsname\relax\def\natexlab#1{#1}\fi
\providecommand{\url}[1]{\href{#1}{#1}}
\providecommand{\dodoi}[1]{doi:~\href{http://doi.org/#1}{\nolinkurl{#1}}}
\providecommand{\doeprint}[1]{\href{http://ascl.net/#1}{\nolinkurl{http://ascl.net/#1}}}
\providecommand{\doarXiv}[1]{\href{https://arxiv.org/abs/#1}{\nolinkurl{https://arxiv.org/abs/#1}}}

\bibitem[{{Barth{\'e}l{\'e}my} {et~al.}(2007){Barth{\'e}l{\'e}my}, {Lilensten},
  \& {Parkinson}}]{barthelemy07}
{Barth{\'e}l{\'e}my}, M., {Lilensten}, J., \& {Parkinson}, C.~D. 2007, \aap,
  474, 301, \dodoi{10.1051/0004-6361:20065891}

\bibitem[{Bell \& Cowan(2018)}]{bell2018}
Bell, T.~J., \& Cowan, N.~B. 2018, The Astrophysical Journal Letters, 857, L20

\bibitem[{{France} {et~al.}(2017){France}, {Fleming}, {West}, {McCandliss},
  {Bolcar}, {Harris}, {Moustakas}, {O'Meara}, {Pascucci}, {Rigby},
  {Schiminovich}, \& {Tumlinson}}]{france2017}
{France}, K., {Fleming}, B., {West}, G., {et~al.} 2017, in Society of
  Photo-Optical Instrumentation Engineers (SPIE) Conference Series, Vol. 10397,
  Society of Photo-Optical Instrumentation Engineers (SPIE) Conference Series,
  1039713, \dodoi{10.1117/12.2272025}

\bibitem[{Koskinen {et~al.}(2022)Koskinen, Lavvas, Huang, Bergsten, Fernandes,
  \& Young}]{koskinen2022mass}
Koskinen, T.~T., Lavvas, P., Huang, C., {et~al.} 2022, The Astrophysical
  Journal, 929, 52

\bibitem[{McCandliss(2003)}]{mccandliss2003molecular}
McCandliss, S.~R. 2003, Publications of the Astronomical Society of the
  Pacific, 115, 651

\bibitem[{Woitke \& Helling(2021)}]{woitke2021ggchem}
Woitke, P., \& Helling, C. 2021, Astrophysics Source Code Library, ascl

\bibitem[{Yelle(2004)}]{yelle2004aeronomy}
Yelle, R.~V. 2004, Icarus, 170, 167

\end{thebibliography}
\bibliographystyle{aasjournal}

\end{document}